# Truly trapped rainbow by utilizing nonreciprocal waveguides

Kexin Liu[1,2] and Sailing He[1,2,*]

1. Department of Electromagnetic Engineering, School of Electrical Engineering, KTH Royal Institute of Technology, Stockholm S-100 44, Sweden

2. Centre for Optical and Electromagnetic Research, State Key Laboratory of Modern Optical Instrumentation, JORCEP (Sino-Swedish Joint Research Center of Photonics), Zhejiang University, Hangzhou 310058, China

*Corresponding author:  sailing@kth.se

**Abstract**

**The concept of a "trapped rainbow" has generated considerable interest for optical data storage and processing. It aims to trap different frequency components of the wave packet at different positions permanently. However, all the previously proposed structures cannot truly achieve this effect, due to the difficulties in suppressing the reflection caused by strong intermodal coupling and distinguishing different frequency components simultaneously. In this article, we found a physical mechanism to achieve a truly "trapped rainbow" storage of electromagnetic wave. We utilize nonreciprocal waveguides under a tapered magnetic field to achieve this and such a trapping effect is stable even under fabrication disorders. We also observe hot spots and relatively long duration time of the trapped wave around critical positions through frequency domain and time domain simulations. The physical mechanism we found has a variety of potential applications ranging from wave harvesting and storage to nonlinearity enhancement.**

Slowing electromagnetic waves is believed to be an attractive technique for enhanced nonlinear optics[1], light harvesting[2,3], and optical signal processing[4,5,6]. Recently, the concept of a "trapped rainbow" has generated considerable interest for potential use in optical data storage and processing[7]. It aims to trap different frequency components of the wave packet at different positions in space permanently[7]. Various waveguide structures are proposed to achieve the trapped rainbow effect, such as tapered waveguides made by negative or hyperbolic metamaterials[7,8,9], tapered plasmonic waveguides,[10,11] and tapered photonic crystal waveguides[12]. These structures can slow down the incident wave around a critical position where the group velocity of the wave is zero[13,14]. However, all these structures cannot truly achieve the trapped rainbow effect. Because of the strong coupling between the forward and backward modes near the critical position, the entire incident wave will be reflected back before reaching the critical position and the wave is not completely standstill in these waveguides[15]. There are other types of structures supposed to completely trapping electromagnetic waves without reflection such as "electromagnetic black holes"[16,17] and magneto-plasmonic waveguides with a block at one end[18,19]. However, these structures are unable to distinguish between different frequency components, because the wave is absorbed or blocked in one singular point for the whole frequency range, rather than trapped at different positions.

In this article, for the first time in the literature, we found a physical mechanism to achieve a truly "trapped rainbow" storage of electromagnetic wave by simultaneously overcoming these two difficulties, namely, suppressing the reflection of the incident wave and distinguishing different frequency components. We utilize a nonreciprocal waveguide under a tapered external magnetic field to achieve this and such a trapping effect is stable even under fabrication disorders. We also observe hot spots and relatively long duration time of the trapped wave around critical positions through frequency domain and time domain simulations with loss taken into consideration. In addition, we investigate the influences of the loss and the applied magnetic field gradient on the trapping effect.

**Physical mechanism.** First, we demonstrate the physical mechanism to achieve trapped rainbow effect by studying the dispersion diagram of the nonreciprocal waveguide and the effect of the applied magnetic field. The waveguide structure is a two-dimensional (2D) slab (infinite in z direction) with three layers (Fig. 1). The top layer is a perfect electric conductor (PEC). The middle layer is a dielectric layer with thickness $d$ and permittivity $\varepsilon_d$. The bottom layer is a gyromagnetic material such as yttrium–iron–garnet (YIG) in microwave frequencies. The permittivity and permeability of the YIG are denoted by $\varepsilon_m$ and $\mu_r$. An external magnetic field $H_0$ is applied on the YIG along the $-z$ direction. In this configuration, $\mu_r$ takes the form[20]

$$\mu_r = \begin{bmatrix} \mu_1 & -i\mu_2 & 0 \\ i\mu_2 & \mu_1 & 0 \\ 0 & 0 & 1 \end{bmatrix} \quad (1)$$

with $\mu_1 = 1 + (\omega_0 + i\nu)\omega_m / [(\omega_0 + i\nu)^2 - \omega^2]$ and $\mu_2 = \omega\omega_m / [(\omega_0 + i\nu)^2 - \omega^2]$, where $\omega_m = 2\pi\gamma H_0$ is the precession frequency, $\nu = \pi\gamma\Delta H$ is determined by the resonance linewidth $\Delta H$, $\omega_m = 2\pi\gamma M_s$ is determined by the saturation magnetization $M_s$ and $\gamma = 2.8\text{E}6$ rad s$^{-1}$G$^{-1}$ is the gyromagnetic ratio. Within the bandgap of the bulk modes in the YIG, we only consider the surface wave of transverse electric (TE) mode ($E_z$, $H_x$, $H_y$) with wave vector $k$ along $x$ in the waveguide[21]. The dispersion relation $\omega(k)$ for this mode is governed by

$$\alpha_d \mu_v + (\alpha_m + \frac{\mu_2}{\mu_1}k)\tanh(\alpha_d d) = 0 \qquad (2)$$

with $\alpha_d = \sqrt{k^2 - \varepsilon_d k_0^2}$, $\alpha_m = \sqrt{k^2 - \varepsilon_m \mu_v k_0^2}$, $\mu_v = \mu_1 - \mu_2^2/\mu_1$ and $k_0 = \omega/c$ (where $c$ is the light speed in vacuum) [22]. The linear term $(\mu_2/\mu_1)k$ with respect to $k$ in equation (2), which originates from the off-diagonal element of $\mu_r$, breaks the symmetry of dispersion relation (i.e., $\omega(k) \neq \omega(-k)$). The propagation of the wave in this waveguide is nonreciprocal. In addition, the dispersion relation is also affected by the thickness of the dielectric layer $d$ [22]. Therefore, we study two lossless cases ($\Delta H = 0$) with very different $d$. In the first case, $d = 0.13\lambda_m$, and the other, $d = 0.013\lambda_m$, where $\lambda_m = 2\pi/k_m$ and $k_m = \omega_m/c$. The other parameters are chosen as follows: $\varepsilon_d = 1$, $\varepsilon_m = 15$, and $M_s = 1780$ G.

For the first case in which $d = 0.13\lambda_m$, $\omega(k)$ changes with $H_0$. Figure 2a shows three curves of $\omega(k)$ with $H_0 = 0.5, 0.6$ and $0.7\, M_s$, respectively. For each curve, there are two frequencies whose group velocity $v_g = d\omega/dk = 0$. One is at the asymptotic frequency $\omega_{-\infty} = \frac{1}{2}\omega_m + \omega_0$ when $k \to -\infty$, and the other one is at a lower frequency $\omega_L$. When $\omega > \omega_{-\infty}$, group velocity $v_g$ points in only one direction and the waveguide is a one-way waveguide. Since $\omega(k)$ can be tuned by $H_0$, we can achieve $v_g = 0$ by controlling $H_0$. Figure 2b shows three relation curves of $k(H_0)$ for $\omega = \omega_m, 1.1\omega_m, 1.2\omega_m$, respectively. For each curve, there are two critical magnetic fields, $H_{c1}$ and $H_{c2}$, where $v_g = 0$. Here, $H_{c1} = (\omega/\omega_m - 1/2)M_s$ and $H_{c2} > H_{c1}$. We analyze the curve with $\omega = 1.2\omega_m$ as an example. For $H_0 < H_{c1}$, the wave has only one finite $k$ and propagates in only one direction with $v_g > 0$ (① in Fig. 2b). For $H_{c1} < H_0 < H_{c2}$, the waveguide has two modes with opposite group velocity directions. For $H_0 > H_{c2}$, the waveguide has no propagating mode. The property of the waveguide around $H_{c1}$ and $H_{c2}$ is very different. For $H_0 \to H_{c1}^+$, the two corresponding modes are separated. One has finite $k$ with $v_g > 0$ (② in Fig. 2b), and the other one has $k \to -\infty$ with $v_g \to 0^-$ (⑤ in Fig. 2b). These two modes cannot couple with each other. For $H_0 \to H_{c2}^-$, the two corresponding modes are nearly degenerated (③ and ④ in Fig. 2b). Although the group velocities of these two modes are opposite, the wave vectors and the modal fields are very close, and consequently these two modes can couple with each other.

Next, we use a tapered $H_0(x)$ to show the physical mechanism for realizing the trapped rainbow in this waveguide (Fig. 2c). Firstly, we assume $H_0(x)$ continuously increases with $x$. $x_{c1}$ and $x_{c2}$ are the corresponding critical positions for $H_{c1}$ and $H_{c2}$ at one certain frequency. The wave is incident at $x < x_{c1}$ along $+x$ direction and it passes through $x_{c1}$ in a unidirectional manner (① in Fig. 2c). The wave continues to propagate without coupling (② in Fig. 2c). When the wave approaches $x_{c2}$ (③ in Fig. 2c), the wave can couple with the corresponding nearly degenerated mode (④ in Fig. 2c). In fact, all the electromagnetic energy can couple back to the -x direction before reaching $x_{c2}$[13]. The wave then travels along the –x direction and approaches $x_{c1}$ where $k \to -\infty$ and $v_g \to 0^-$ (⑤ in Fig. 2c). There is no coupling between ② and ⑤ in Fig. 2c and the wave cannot penetrate $x_{c1}$, so the wave is trapped at $x_{c1}$. Secondly, we consider some disorders in the waveguide such as surface roughness or a non-homogeneous material. The disorders at $x < x_{c1}$ will not generate backscattering, since the waveguide is a one-way waveguide in this region[20]. The disorders at $x_{c1} < x < x_{c2}$ can generate a scattering wave. The scattering wave traveling in the –x direction will approach $x_{c1}$ and be trapped there, while the scattering wave traveling in the +x direction will couple back to the –x direction and still become trapped at $x_{c1}$. Note that, at the frequencies within the bandgap of the bulk mode in YIG, the scattering cannot propagate inside the YIG layer. Thirdly, the critical positions are different for different frequencies (Fig. 2b), so the different frequency components of a wave packet can be trapped at different positions. To summarize, this waveguide can cage the wave between the two critical positions and achieve the trapped rainbow effect.

For the second case in which $d = 0.013\lambda_m$, $\omega(k)$ also changes with $H_0$. Figure 2d shows three curves of $\omega(k)$ with $H_0 = 0.5, 0.6$ and $0.7\, M_s$, respectively. For each curve, group velocity $v_g$ points in only one direction and $v_g \to 0^+$ at the asymptotic frequency $\omega_{-\infty} = \frac{1}{2}\omega_m + \omega_0$ when $k \to -\infty$. Figure 2e shows the relation curves of $k(H_0)$ for $\omega = \omega_m, 1.1\omega_m$ and $1.2\omega_m$, respectively. The critical magnetic field is $H_c = (\omega/\omega_m - 1/2)M_s$, where $v_g = 0$. We analyze the curve with $\omega = 1.2\omega_m$ as an example. For $H_0 < H_c$, the wave has only one finite $k$ and propagates in only one direction with $v_g > 0$ (① and ② in Fig. 2e). When $H_0 \to H_c^-$, the wave has $k \to -\infty$ and $v_g \to 0^+$ (③ Fig. 2e). For $H_0 > H_c$, there is no propagating mode. We next use a continuously increasing $H_0(x)$ with x to achieve the trapped rainbow effect. The physical mechanism is quite simple in this case: Set $x_c$ as the corresponding critical position for $H_c$ at one certain frequency. The wave is incident at $x < x_c$ along +x direction (① in Fig. 2f). Since there is no mode propagating in the -x direction, the coupling does not exist. Disorders cannot generate backscattering waves in this one-way waveguide or propagating waves in the YIG layer. Therefore, the wave continues to propagate along +x with decreasing $v_g$ (② in Fig. 2f). Finally, the wave approaches $x_c$ with $v_g \to 0^+$. Since it cannot penetrate $x_c$, the wave is trapped at $x_c$ (③ in Fig. 2f). As critical position $x_c$ is related to the frequency, different frequency components of a wave packet can be trapped at different positions. Therefore, this waveguide can achieve the trapped rainbow effect by trapping the wave at the critical position.

**Trapped rainbow.** We simulate the trapped rainbow effect in the nonreciprocal waveguide structures. Both frequency domain and time domain simulations for two cases $d = 0.013\lambda_m$ and $d = 0.13\lambda_m$ are conducted, taking loss ($\Delta H \neq 0$) into consideration. According to the analysis above, a continuously increasing $H_0(x)$ with $x$ can trap the wave. We choose $H_0(x)$ changing linearly with $x$:

$$H_0(x) = (\alpha x/\lambda_m + 0.4)M_s \tag{3}$$

where $\alpha$ indicates the increasing rate of $H_0$.

For the first case $d = 0.13\lambda_m$, we initially conduct the frequency domain simulation by the finite element method (COMSOL). The parameters are chosen as follows: $\alpha = 0.1$ and $\Delta H = 2$ Oe. The matched field is excited at $x = 0$. The amplitude $E_z$ distributions for $\omega = \omega_m$, $1.1\omega_m$ and $1.2\omega_m$ (Fig. 3 a-c) show that the field is well confined at the dielectric-YIG interface ($y = 0$) and enhanced in a region between $x_{c1}$ and $x_{c2}$. Here $x_{c1} = \lambda_m$, $2\lambda_m$ and $3\lambda_m$, and $x_{c2} = 2.03\lambda_m$, $2.99\lambda_m$ and $3.95\lambda_m$ for $\omega = \omega_m$, $1.1\omega_m$ and $1.2\omega_m$, respectively. Fig. 3d shows the normalized amplitude of $E_z$ along the interface. The amplitude has no ripples on the left side of $x_{c1}$, since the wave propagates only one-way in this region. On the right side of $x_{c1}$, the ripples of the amplitude demonstrate an interface between the waves propagating in $\pm x$ direction. The wave propagating in the $-x$ direction comes from the coupling when the wave in the $+x$ direction approaches $x_{c2}$. In addition, when the wave approaches $x_{c2}$, the group velocity is slower, so the field is also enhanced in this process. Therefore, around these two positions, we can observe hot spots. Between these two positions, the intensity of the field oscillates and has several peaks.

We then study the propagation of a wave packet in the time domain for case $d = 0.13\lambda_m$ (see Supplementary Information for details of the calculation method). A Gaussian wave packet with center frequency $\omega_c = 1.1\omega_m$ is injected into the waveguide with $\alpha = 0.1$ and $\Delta H = 2$ Oe. Figure 3e-i shows the $E_z$ amplitude distribution of the wave packet at different times (see also Supplementary Movie 1). Figure 3j shows the nomalized distribution of $E_z$ amplitude along the dielectric-YIG interface at different times. They show that the wave packet initially propagates in $+x$ direction and is compressed and enhanced when approaching $x_{c2} = 2.99\lambda_m$. The wave then goes back and antenuates gradually at $x_{c1}=2\lambda_m$. Figure 3k shows the time evolution of the $E_z$ amplitude at four positions: $x = 0.50\lambda_m$ (blue), $2.87\lambda_m$ (green), $2.58\lambda_m$ (red) and $2.00\lambda_m$ (black) on the interface. Note that between $x_{c1}$ and $x_{c2}$, such as $x = 2.58\lambda_m$, the field can have two peaks in the time domain.

For the second case $d = 0.013\lambda_m$, we first simulate the propagation of the wave in the frequency domain. The parameters are chosen as follows: $α=0.2$ and $ΔH = 1$ Oe. The matched field is excited at $x = 0$. The amplitude $E_z$ distributions for $ω = ω_m$, $1.1ω_m$ and $1.2ω_m$ (Fig. 4a-c) show that the field is well confined at the dielectric-YIG interface ($y = 0$) and enhanced around the corresponding critical positions $x_c = 0.5\lambda_m$, $1.0\lambda_m$ and $1.5\lambda_m$, respectively. Figure 4d shows the normalized amplitude of $E_z$ along the interface. The amplitude increases monotonically from $x = 0$ to the critical positions, which demonstrates that the wave is trapped and extremely enhanced at the critical positions.

Next, we study the propagation of a wave packet in the time domain for $d = 0.013\lambda_m$. A Gaussian wave packet with center frequency $ω_c = 1.1ω_m$ is injected into the waveguide with $α=0.2$ and $ΔH = 3$ Oe. Figures 4e-i show the $E_z$ amplitude distributions of the wave packet at different times (see also Supplementary Movie 2). Figure 4j shows the nomorlized distribution of $E_z$ amplitude along the dielectric-YIG interface at different times. They show that the wave packet is compressed, enhanced, and trapped around the critical position, and then antenuates gradually. Figure 4k shows the time evolution of the $E_z$ amplitude at $x = 0.500\lambda_m$ (blue), $0.935\lambda_m$ (green), $1.000\lambda_m$ (red) and $1.050\lambda_m$ (black). We notice that the amplitude of the wave packet achieves the maximum value before it reaches the critical position. This means that, although slower group velocity can enhance the field, when the wave packet approaches the critical position, the loss increases substantially and the intensity of the wave decreases. We also notice that the wave packet can remain at the critical position in for a relatively long time. Define the duration time $Δt$ as the time period when the amplitude of $E_z$ is greater than half of the maximum value. Figure 5l shows $Δt$ for three cases: (1) $α = 0.2$, $ΔH = 3$ Oe; (2) $α = 0.2$, $ΔH = 2$ Oe and (3) $α = 0.3$, $ΔH = 3$ Oe. It demonstrates that $Δt$ has maximum value at the critical position. In addition, lower $ΔH$ gives less loss and contributes to longer duration, and $α$ seems not to affect the duration time.

We next investigate the trapped rainbow effect in the waveguides with disorders. For $d = 0.13\lambda_m$, two dielectric slabs (white slabs in Fig. 5a) with permittivity $ε =10$, width $Δx=\lambda_m/20$ and height $Δy=\lambda_m/20$ are inserted at $x = \lambda_m$ and $x = 1.5\lambda_m$ through the dielectric-YIG interface. The other parameters are the same as those in Fig. 3b. The 2D distribution of $E_z$ (Fig. 5a) and the amplitude $E_z$ along the dielectric-YIG interface (Fig. 5b) are almost identical to the results in Fig. 3b and 3d. This demonstrates that the waveguide can preserve the trapped rainbow effect under disorders. For $d = 0.013\lambda_m$, one dielectric slab (white slab in Fig. 5c) with permittivity $\varepsilon =10$, width $\Delta x=\lambda_m/50$ and height $\Delta y=\lambda_m/100$ is inserted at $x = 0.5\lambda_m$ through the dielectric-YIG interface. The other parameters are the same as those in Fig. 4b. The 2D distribution of $E_z$ (Fig. 5c) and the amplitude $E_z$ along the dielectric-YIG interface (Fig. 5d) are almost identical to the results in Fig. 4b and 4d. This demonstrates that this waveguide can also preserve the trapped rainbow effect under disorders. Compared with other slow-light structures, in which the disorder can generates reflection and destroys the trapped rainbow effect (see Supplementary Information), the nonreciprocal waveguides are immune to the disorders, which is an important advantage for achieving the trapped rainbow effect.

Finally, to get much stronger enhancement of the electromagnetic field, we investigate the influence of $\alpha$ and $\Delta H$ on the enhancement at $\omega = 1.1\omega_m$ for $d = 0.013\lambda_m$ in the frequency domain. We define an enhancement factor (EF) as ratio of the amplitude of $E_z$ at the critical position to the amplitude at *x*=0 on the dielectric-YIG interface. We sweep $\alpha$ from 0.13 to 0.40 and sweep $\Delta H$ from 1 Oe to 10 Oe to find the corresponding EF (Fig. 6a). Figure 6b shows that, for a fix $\alpha = 0.2$, EF decreases as $\Delta H$ increases, because larger $\Delta H$ contributes more loss. Figure 6c shows that, for a fix $\Delta H = 3$ Oe, EF increases with $\alpha$. This is because in this one-way waveguide, the more rapidly increasing rate will contribute a shorter way to reach the critical position with less loss and higher EF, rather than generate reflection. Figure 6a also shows the contour lines for EF = 1 and EF = 10. To summarize, smaller $\Delta H$ and larger $\alpha$ can contribute to a higher EF.

In conclusion, we found a physical mechanism to achieve a truly "trapped rainbow" storage of waves by utilizing nonreciprocal waveguides. Two nonreciprocal waveguides with different thicknesses under tapered applied magnetic fields are investigated via frequency domain and time domain simulations. Results demonstrate that both can achieve the trapped rainbow effect even under disorders. One can cage the wave between two critical positions, and the other can trap the wave at one critical position. The field can be enhanced at the trapped positions. In addition, low loss is essential to achieve strong enhancement of the field and long duration time, and a more rapidly increasing rate of the tapered external magnetic field can produce stronger enhancement of the field. To verify the trapped rainbow effect in future microwave experiments, a very low loss YIG film is necessary and the tapered external magnetic field should have high gradients. Further research may make use of the physical mechanism proposed here to investigate light wave or acoustic wave trapping[23]. Applications ranging from wave harvesting and storage to nonlinearity enhancement might also benefit from the physical mechanism we suggest.

**Acknowledgements**

We thank A. Tork for his helpful discussion. K. Liu thanks China Scholarship Council (CSC) (201406320056). Support for this work was provided by Swedish VR grant (621-2011-4620); AOARD (FA2386-14-1-0026); National Natural Science Foundation of China (NSFC) (61178062, 91233208).

**Author contributions**

S. He conceived the ideas and supervised the research. K. Liu performed the calculations and simulations. S. He and K. Liu made the theoretical analysis and wrote the paper.

**Competing financial interests**

The authors declare no competing financial interests.

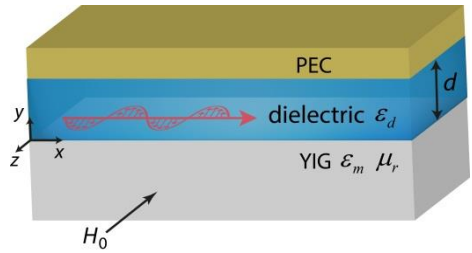

**Figure 1 | Waveguide structure.** The waveguide structure is a slab with three layers. The top layer is a perfect electric conductor (PEC). The middle layer is a dielectric layer with thickness $d$ and permittivity $\varepsilon_d$. The bottom layer is an yttrium–iron–garnet (YIG) layer, which is gyromagnetic material in microwave frequencies. The permittivity and permeability of the YIG are denoted by $\varepsilon_m$ amd $\mu_r$. An external magnetic field $H_0$ is applied on the YIG along the –z direction. The wave of transverse electric (TE) mode ($E_z$, $H_x$, $H_y$) goes along $x$ direction.

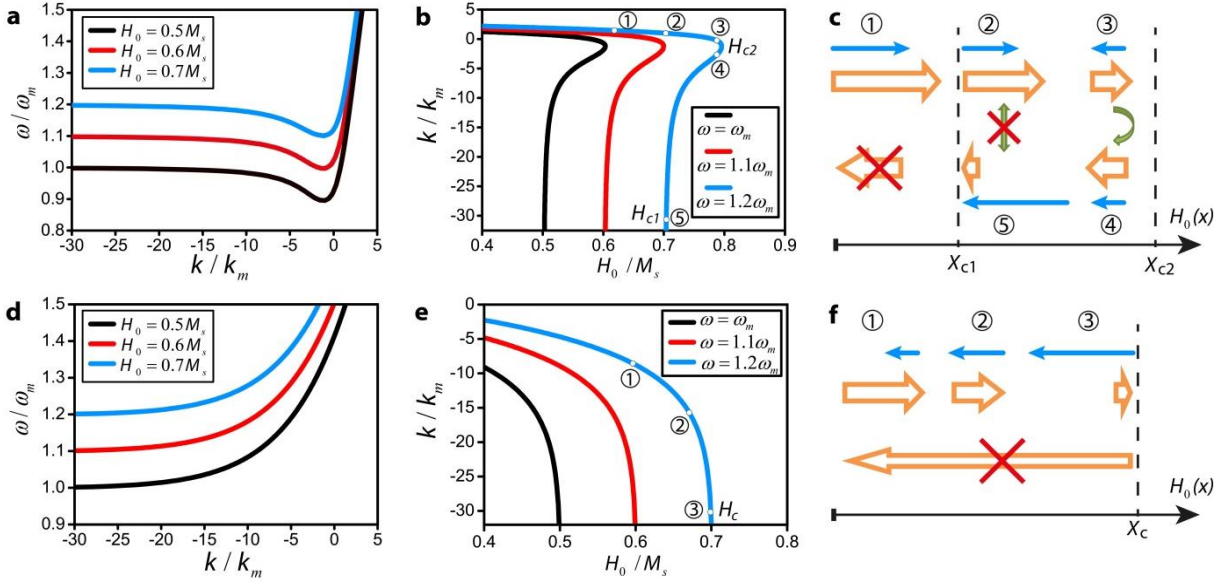

**Figure 2 | Physical mechanism of trapped rainbow in nonreciprocal waveguide. a,** Dispersion relation for $d = 0.13\lambda_m$, $H_0 = 0.5M_s$ (black), $0.6M_s$ (red) and $0.7M_s$ (blue). **b,** Relation curve of $k(H_0)$ for $d = 0.13\lambda_m$, $\omega = \omega_m$ (black), $1.1\omega_m$ (red) and $1.2\omega_m$ (blue). $H_{c1}$ and $H_{c2}$ indicate the critical fields for $\omega = 1.2\omega_m$. ① ② ③ ④ ⑤ denote five positions (white dots) on the relation curve for $\omega = 1.2\omega_m$. At ① $H_0 < H_{c1}$, at ② and ⑤ $H_0 \to H_{c1}^+$, and at ③ and ④ $H_0 \to H_{c2}^-$. **c,** Schematic diagram for $v_g$ (yellow wide arrow) and $k$ (blue thin arrow) at corresponding positions marked in **b**. $H_0(x)$ continuously increases with $x$. $x_{c1}$ and $x_{c2}$ denote the corresponding critical positions for $H_{c1}$ and $H_{c2}$. ① denotes the incident wave at $x < x_{c1}$. The yellow arrow with a red cross denotes that the wave cannot propagate in $-x$ direction at $x < x_{c1}$. The green straight arrow with a red cross denotes no coupling between the waves at ② and ⑤. The green curved arrow denotes coupling between the waves at ③ and ④. **d,** Dispersion relation for $d = 0.013\lambda_m$, $H_0 = 0.5M_s$ (black), $0.6M_s$ (red) and $0.7M_s$ (blue). **e,** Relation curve of $k(H_0)$ for $d = 0.013\lambda_m$, $\omega = \omega_m$ (black), $1.1\omega_m$ (red) and $1.2\omega_m$ (blue). $H_c$ indicates the critical field for $\omega = 1.2\omega_m$. ① ② ③ denote three positions (white dots) on the relation curve for $\omega = 1.2\omega_m$. **f,** Schematic diagram for $v_g$ (yellow wide arrow) and $k$ (blue thin arrow) at the corresponding positions marked in **e**. $H_0(x)$ continuously increases with $x$. $x_c$ denotes the corresponding critical position for $H_c$. The yellow wide arrow with a red cross denotes that the wave cannot propagate in $-x$ direction. The other parameters are as follows: $\varepsilon_d = 1$, $\varepsilon_m = 15$ and $M_s = 1780$ G.

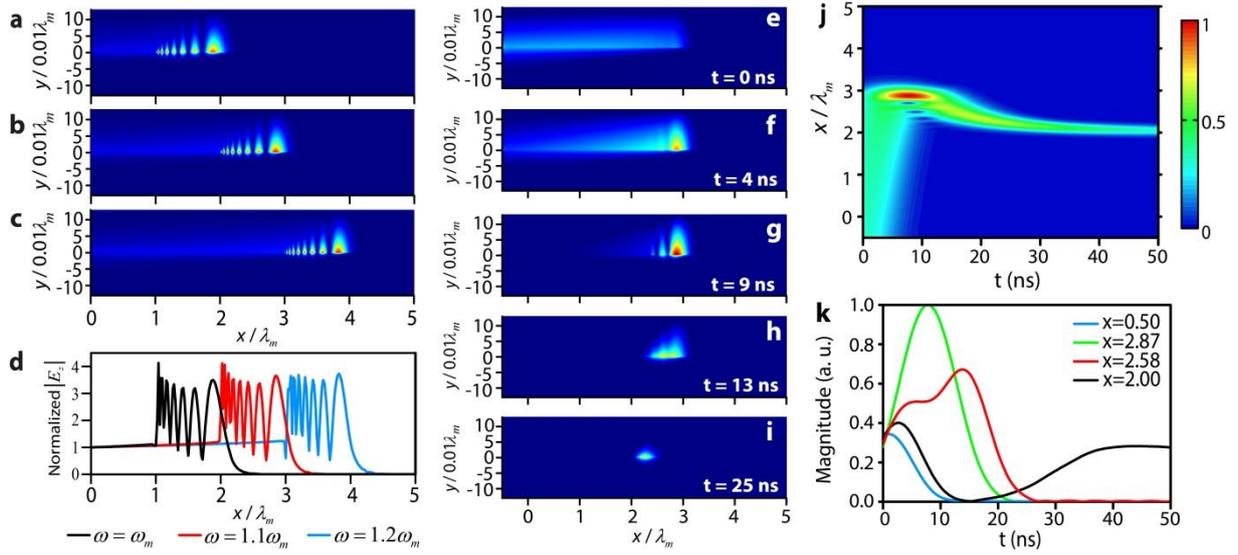

**Figure 3 | Simulated trapped rainbow for $d = 0.13\lambda_m$.** In the frequency domain, matched fields are excited at *x*=0 in the waveguide with $\alpha = 0.1$ and $\Delta H = 2$ Oe. The distribution of the $E_z$ amplitude is plotted for **a** $\omega = \omega_m$, **b** $1.1\omega_m$ and **c** $1.2\omega_m$, respectively. **d,** $E_z$ amplitude along the dielectric-YIG interface for $\omega = \omega_m$ (black), $1.1\omega_m$ (red) and $1.2\omega_m$ (blue). In the time domain, a Gaussian wave packet with center frequency $\omega_c = 1.1\omega_m$ is injected into the waveguide with $\alpha = 0.1$ and $\Delta H = 2$ Oe. The $E_z$ amplitude distribution of the wave packet is plotted at **e** 0 ns, **f** 4 ns, **g** 9ns, **h** 13 ns, and **i** 25 ns, respectively. **j,** Normalized distribution of the $E_z$ amplitude along the dielectric-YIG interface at different time. **k,** Time evolution of the $E_z$ amplitude at $x = 0.50\lambda_m$ (blue), $2.87\lambda_m$ (green), $2.58\lambda_m$ (red) and $2.00\lambda_m$ (black). The other parameters are $\varepsilon_d = 1$, $\varepsilon_m = 15$ and $M_s = 1780$ G.

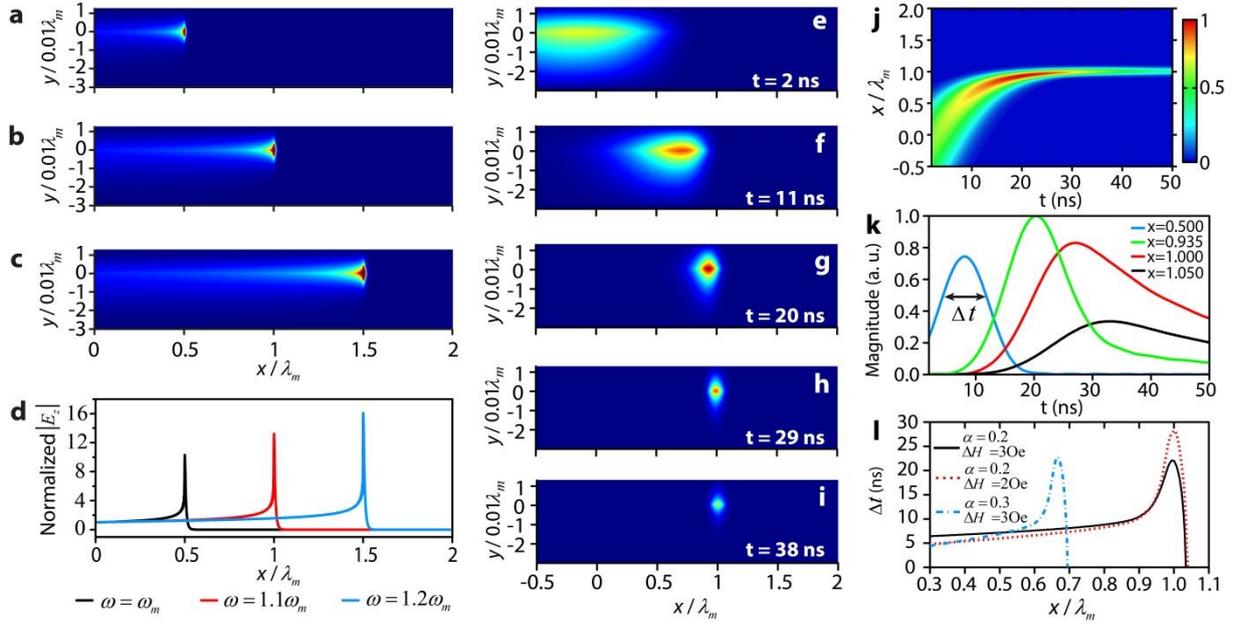

**Figure 4 | Simulated trapped rainbow for $d = 0.013\lambda_m$.** In the frequency domain, matched fields are excited at $x=0$ in the waveguide with $\alpha = 0.2$ and $\Delta H = 1$ Oe. The distribution of the $E_z$ amplitude is plotted for **a** $\omega = \omega_m$, **b** $1.1\omega_m$ and **c** $1.2\omega_m$, respectively. **d,** $E_z$ amplitude along the dielectric-YIG interface for $\omega = \omega_m$ (black), $1.1\omega_m$ (red) and $1.2\omega_m$ (blue). In the time domain, a Gaussian wave packet with center frequency $\omega_c = 1.1\omega_m$ is injected into the waveguide with $\alpha = 0.2$ and $\Delta H = 3$ Oe. The $E_z$ amplitude distribution of the wave packet is plotted at **e** 2 ns, **f** 11 ns, **g** 20 ns, **h** 29 ns, and **i** 38 ns, respectively. **j,** Normalized distribution of the $E_z$ amplitude along the dielectric-YIG interface at different times. **k,** Time evolution of the $E_z$ amplitude at $x = 0.500\lambda_m$ (blue), $0.935\lambda_m$ (green), $1.000\lambda_m$ (red) and $1.050\lambda_m$ (black). **l,** Duration time $\Delta t$ for three cases: (1) $\alpha = 0.2$ and $\Delta H = 3$ Oe (black solid line), (2) $\alpha = 0.2$ and $\Delta H = 2$ Oe (red dot line), and (3) $\alpha = 0.3$ and $\Delta H = 3$ Oe (blue dot-dash line). The other parameters are as follows: $\varepsilon_d = 1$, $\varepsilon_m = 15$ and $M_s = 1780$ G.

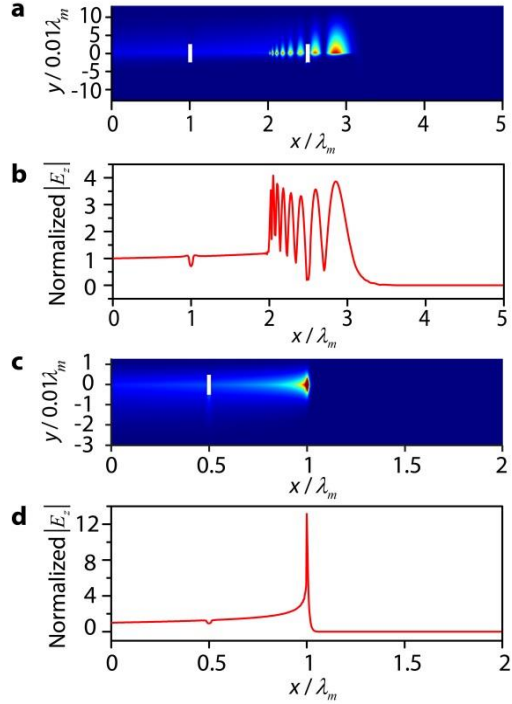

**Figure 5 | Trapped rainbow effect under disorders in the nonreciprocal waveguides.** For $d = 0.13\lambda_m$, two dielectric slabs (white slabs) with permittivity $\varepsilon =10$, width $\Delta x=\lambda_m/20$ and height $\Delta y=\lambda_m/20$ are inserted at $x = \lambda_m$ and $x = 1.5\lambda_m$ through the dielectric-YIG interface. The other parameters are the same as those in Fig. 3b. **a,** Distribution of the $E_z$ amplitude. **b,** $E_z$ amplitude along the dielectric-YIG interface. For $d = 0.013\lambda_m$, one dielectric slabs (white slabs) with permittivity $\varepsilon =10$, width $\Delta x=\lambda_m/50$ and height $\Delta y=\lambda_m/100$ is inserted at $x = 0.5\lambda_m$ through the dielectric-YIG interface. The other parameters are the same as those in Fig. 4b **c,** Distribution of the $E_z$ amplitude. **d,** $E_z$ amplitude along the dielectric-YIG interface.

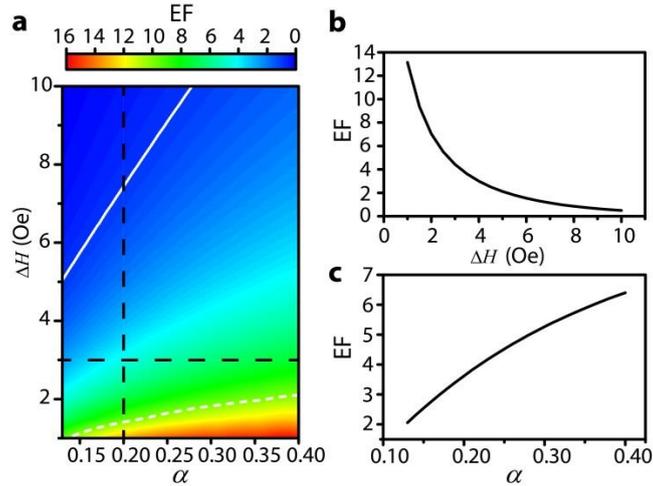

**Figure 6 | The influence of $\alpha$ and $\Delta H$ on the enhancement factor (EF).** **a,** EF changes with $\alpha$ from 0.13 to 0.4 and $\Delta H$ from 1 Oe to 10 Oe. The white solid (dashed) line is the contour line for EF=1 (EF=10). **b,** EF changes with $\Delta H$ for a fix $\alpha = 0.2$, corresponding to the vertical black dash line in **a**. **c,** EF changes with $\alpha$ for a fix $\Delta H$, corresponding to horizontal black dash line in **a**. The other parameters are as follows: $\omega = 1.1\omega_m$, $d = 0.013\lambda_m$, $\varepsilon_d = 1$ and $\varepsilon_m = 15$, $M_s = 1780$ G.

# Supplementary Information

## Truly trapped rainbow by utilizing nonreciprocal waveguides

Kexin Liu[1,2] and Sailing He[1,2,*]

*1. Department of Electromagnetic Engineering, School of Electrical Engineering, KTH Royal Institute of Technology, Stockholm S-100 44, Sweden*

*2. Centre for Optical and Electromagnetic Research, State Key Laboratory of Modern Optical Instrumentation, JORCEP (Sino-Swedish Joint Research Center of Photonics), Zhejiang University, Hangzhou 310058, China*

*Corresponding author: sailing@kth.se*

**A. Method for time domain simulation**

The method for the time domain simulation is the same as the method in [1]. In our paper, the propagation of the wave packet in the time domain is formed by the superposition of 51 modulated frequency components with a central frequency $\omega_c = 1.1\omega_m$. In a frequency domain simulation each frequency component $\psi(\omega_n)$ is obtained by exciting a 1A line current at $x = -0.5\lambda_m$ in COMSOL. The frequency sequence $\omega_n$ is an arithmetic sequence from $1.08\omega_m$ to $1.12\omega_m$ with a common difference of $\Delta\omega = \omega_m/1250$. The amplitude of each component is modulated by a Gaussian function $A(\omega_n) = \exp[-\frac{1}{200}\left(\frac{\omega_n-\omega_c}{\Delta\omega}\right)^2]$ as shown in Fig. S1. The superposition of the modulated components gives the propagation of the wave packet $\Psi$ in the time domain by the following formula

$$\Psi = \sum_n A(\omega_n)\,\psi(\omega_n)\exp(i\omega_n t) \tag{S1}$$

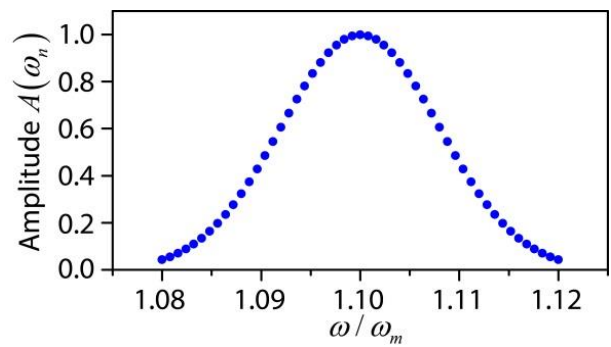

**Figure S1 | Gaussian function $A(\omega_n)$.** 51 frequency components are modulated by the Gaussian function $A(\omega_n)$ to construct the propagation of the wave packet in the time domain.

## B. Reflection caused by disorders in reciprocal waveguides

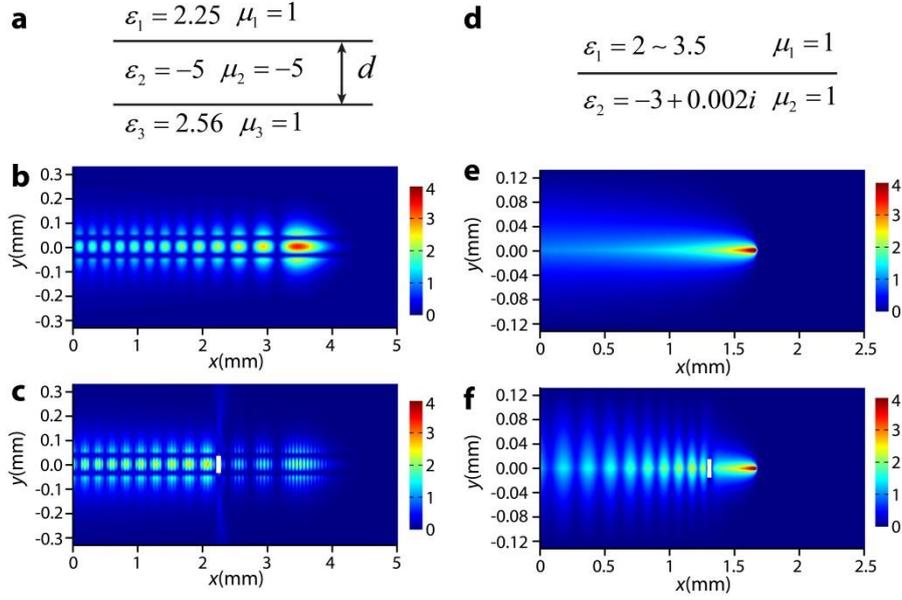

**Figure S2 | Reflection caused by disorders in reciprocal waveguides. a,** Structure of a three-layer tapered waveguide. The bottom and upper layers are dielectric materials with $\varepsilon_1 = 2.25$, $\mu_1 = 1$ and $\varepsilon_3 = 2.56$, $\mu_3 = 1$. The middle layer is double negative metamaterial with $\varepsilon_2 = -5$ and $\mu_2 = -5$. The thickness $d$ of the middle layer changes linearly from 55um to 51.5um with $x$ from 0mm to 5mm. The frequency is 1THz and the critical thickness is 52.45um. **b,** Distribution of $|H_z|$ without disorders for the waveguide in **a**. **c,** Distribution of $|H_z|$ with a dielectric slab (white rectangle) with $\varepsilon_r = 10$, $\mu_r = 1$ and $\Delta x = \Delta y = 0.7$mm inserted in the waveguide in **a**. **d,** Structure of a two-layer waveguide. The upper layer is a dielectric layer with $\varepsilon_1$ changing from 2 to 3.5 with $x$ from 0mm to 2.5mm. The bottom layer has negative permittivity $\varepsilon_2 = -3 + 0.002i$. The operation frequency is 0.3GHz and the critical value of $\varepsilon_1$ is -3. **e,** Distribution of $|H_z|$ without disorders for the waveguide in **d**. **f,** Distribution of $|H_z|$ with a dielectric slab (white rectangle) with $\varepsilon_r = 10$, $\mu_r = 1$ and $\Delta x = \Delta y = 0.03$mm inserted in the waveguide in **d**.

Compared to the nonreciprocal waveguides suggested in our paper, reciprocal waveguides are sensitive to disorders. We study two slow wave structures to show the influence of the disorders. The first one is a three-layer tapered waveguide with the same material parameters as those in [1, 2] (Fig. S2a). The bottom and upper layers are dielectric materials with $\varepsilon_1 = 2.25$, $\mu_1 = 1$ and $\varepsilon_3 = 2.56$, $\mu_3 = 1$. The middle layer is double negative metamaterial with $\varepsilon_2 = -5$ and $\mu_2 = -5$. The thickness $d$ of the middle layer changes linearly from 55um to 51.5um with $x$ from 0 mm to 5mm. The frequency is 1THz and the critical thickness is 52.45um where the group velocity $v_g$ is zero. Figure S2b shows the distribution of $|H_z|$ without disorders. Figure S2c shows the distribution of $|H_z|$, when we insert a dielectric slab (white rectangle) with $\varepsilon_r = 10$, $\mu_r = 1$ and $\Delta x = \Delta y = 0.7$mm. The disorder generates direct reflection in this waveguide. The second is a two-layer structure supporting the surface wave (Fig. S2d). The upper layer is a dielectric layer with $\varepsilon_1$ changing from 2 to 3.5 with *x* from 0mm to 2.5mm. The bottom layer has negative permittivity $\varepsilon_2 = -3 + 0.002i$. The operation frequency is 0.3GHz and the critical value of $\varepsilon_1$ is -3 where $v_g \to 0$ and the wave vector $k \to \infty$. Figure S2e shows the distribution of $|H_z|$ without disorders. Figure S2f shows the distribution of $|H_z|$, when we insert a dielectric slab (white rectangle) with $\varepsilon_r = 10$, $\mu_r = 1$ and $\Delta x = \Delta y = 0.03$mm. The disorder can also generate direct reflection in this waveguide.